\documentclass[prb,preprint,showpacs]{revtex4}

\usepackage{amsfonts}
\usepackage{amssymb}
\usepackage{amsmath}
\usepackage[english]{babel}
\usepackage{graphicx}
\usepackage{pslatex}

\begin{document}

\author{A. Castro}
\email{alberto@physik.fu-berlin.de}
\affiliation{Institut f{\"u}r Theoretische Physik and European Theoretical Spectroscopy Facility,
Freie Universit{\"a}t Berlin, Arnimallee 14, D-14195 Berlin, Germany}

\author{E. R{\"a}s{\"a}nen}
\email{erasanen@jyu.fi}
\affiliation{Nanoscience Center, Department of Physics, 
University of Jyv{\"a}skyl{\"a}, FI-40014 Jyv{\"a}skyl{\"a}, Finland}

\author{C. A. Rozzi}
\email{rozzi@unimore.it}
\affiliation{INFM-CNR National Research Center "S3", via Campi 213a, I-41100 Modena, Italy}

\title{Exact Coulomb cutoff technique for supercell calculations in two dimensions}

\pacs{71.15.Mb, 73.21.La, 71.10.Ca}


\begin{abstract}

  We present a reciprocal space technique for the calculation of the
  Coulomb integral in two dimensions in systems with reduced
  periodicity, i.e., finite systems, or systems that are periodic only
  in one dimension. The technique consists in cutting off the
  long-range part of the interaction by modifying the expression for
  the Coulomb operator in reciprocal space. The physical result
  amounts in an effective screening of the spurious interactions
  originated by the presence of ghost periodic replicas of the
  system. This work extends a previous report [C. A. Rozzi {\em et
    al}, Phys. Rev. B {\bf 73}, 205119 (2006)], where
  three-dimensional systems were considered. We show that the use of
  the cutoffs dramatically enhances the accuracy of the calculations,
  and it allows to describe two-dimensional systems of reduced
  periodicity with substantially less computational effort. In
  particular, we consider quantum-dot arrays having potential
  applications in quantum information technology.

\end{abstract}

\maketitle

\section{Introduction}

The interest in low-dimensional electronic structures has increased steadily
during the past few decades. This is mostly due to breakthroughs
in semiconductor technology in the 1970s and early 1980s. At present, 
low-dimensional systems form a significant portion of the whole field of 
condensed matter physics. Some examples are layered semiconductor devices
such as metal-oxide-semiconductor field effect transistors, quantum
Hall systems, spintronic devices, and quantum wells, wires, 
and dots~\cite{qd_reviews} (QDs).

Following the advances in constructing techniques for single two-dimensional (2D) 
QDs with tunable atom-like properties, it has become possible to {\em couple}
QDs to form artificial ``molecules''. Coupled QDs have significant
potential for solid-state quantum computation through, e.g., coherent
manipulation of spins.~\cite{qd_spin} Furthermore, extended lattices
or arrays of QDs have been fabricated,~\cite{kouwenhoven} which,
in addition to the proposed applications in quantum information,~\cite{shiraishi} 
show interesting magnetic phase transitions~\cite{koskinen,karkkainen} 
which may be exploited in quantum transport and spintronics.

From the theoretical point of view, dealing with systems of arbitrary
periodic dimensionality may be complicated.  In the simplest case of
a fully periodic lattice of elements, periodic boundary
conditions are applied at every cell border, and Bloch's theorem
describes the discrete-translation invariant form of the
orbitals. On the other hand, in all the cases in which the system has
reduced periodicity, the use of a supercell with periodic boundary
conditions in all the directions becomes problematic. In fact, the
response function of a periodic lattice is generally very different
from the response of a system with reduced periodicity (such as an
isolated system, a chain, a slab, etc.) and the convergence of the
fully periodic quantity to the reduced-periodic ones as a function of
the supercell size is very slow.  A large supercell is a numerical
disadvantage, but it is specially necessary to avoid the influence of
the periodic images if long range operators are used.

The main issue here is indeed the computation of a long-range
operator, i.e., the Hartree (or ``Coulomb'') potential:
\begin{equation}\label{HertreeInt}
V[n](\mathbf{r}) = \int\! {\rm d}^3 r'
\frac{n(\mathbf{r}')}{\vert \mathbf{r}-\mathbf{r}'\vert}\,,
\end{equation}
which is ubiquitous in Science, and which we study here in
the context of 2D electronic structure calculations. In particular,
we will exemplify our approach by utilizing density-functional theory
(DFT), although the method we propose can also be useful in different fields. In
Fourier space, the convolution integral (\ref{HertreeInt}) is transformed into a trivial product.
This fact adds to the other undoubted advantages of the supercell approach,
such as the natural inclusion of the periodic boundary conditions, and
the existence of very efficient fast Fourier transform algorithms.

The attempt to retain these advantages, i.e., to compute the Hartree
integral in reciprocal space, has led to the creation of several
cutoff schemes for finite systems, whose main intent is to provide an
effective truncated Coulomb interaction such that the system becomes
unaware of the existence of its periodic
replicas.\cite{Makov95,Jarvis97,Castro03} More recently, an exact
scheme was proposed to achieve a broader goal,\cite{rozzi06} namely to
truncate the Coulomb interaction in a 3D system in the
dimensions along which the system is confined, leaving it 
long-ranged in the dimensions in which the system is periodic.

In this paper we show that a similar scheme can be drawn in a 2D
space, allowing us to correct the spurious supercell effect when
treating finite systems and one-dimensional (1D) chains. In particular
we focus on the case of a single infinite chain of 2D few-electron
QDs, that, in a classical supercell approach, would mistakenly
appear as periodic in both directions, while it should be
treated as a truly periodic system only along the x direction.  We
study the magnetic ground state of this system, and show that the use
of the cutoff allows to speedup the calculations of ground state
fundamental quantities such as the Fermi level. The technique is exact
if the computation parameters are chosen judiciously. The possibility
of isolating the chain from the replicas permits to get an insight
about the role of inter-chain interaction in determining the conductive 
of insulating character of the chain.



\section{Method}

We follow closely the procedure described in Ref.~[\onlinecite{rozzi06}]. 
The Hartree integral is
the convolution of the charge density $n$ with the Coulomb
interaction potential $v$. In 2D,
\begin{align}\label{defHartree}
  V(x,y)&=
  \iint_{\mathrm{space}}
  \frac{n(x',y')}{\sqrt{(x-x')^2+(y-y')^2}}
  \mathrm{d}x'\mathrm{d}y' \nonumber \\
  &=\iint_{\mathrm{space}}
  n(x',y')v(\mathbf{r}-\mathbf{r}')
  \mathrm{d}x'\mathrm{d}y', 
\end{align}
We consider the charge density to be in a unit cell. This unit cell
may then be replicated in both directions to fill all 2D space (the
two dimensional periodic case, or ``2D/2D''), replicated in only one
direction (``2D/1D''), or not replicated at all (finite case, or
``2D/0D'').  If we move to reciprocal space, the unit cell is always
replicated periodically in both directions, which is undesired in both
the 2D/1D and 2D/0D cases.  The Hartree integral in reciprocal space
reduces to the simple multiplication:
\begin{equation}\label{defHartreeReciproc}
  V(\mathbf{G})=
  n(\mathbf{G})v(\mathbf{G}),
\end{equation}
where $\mathbf{G}=(G_x,G_y)$ are the reciprocal vectors. The Fourier transform
of $v(x,y)$ can be readily computed:
\begin{equation}
\label{eq:2d-2d}
  v(G_x, G_y)=\int_{-\infty}^{+\infty}\!\!\!\!\!\!\mathrm{d}x
  \int_{-\infty}^{+\infty}\!\!\!\!\!\!\mathrm{d}y
  \frac{\mathrm{e}^{i(G_x x+G_y y)}}{\sqrt{x^2+y^2}} 
=\frac{2\pi}{G}.
\end{equation}
This expression implies full 2D periodicity, and therefore it contains
spurious terms if the periodicity is reduced.
Our aim is to modify the expression of the Coulomb interaction 
$v(\mathbf{G})\to\tilde v(\mathbf{G})$
to an effectively truncated interaction:
\begin{equation}
  \tilde v(\mathbf{r}) =
  \begin{cases}
    \frac{1}{r}&    \text{if $\mathbf{r}\in\cal{D}$} \\
    0&          \text{if $\mathbf{r}\notin\cal{D}$},
  \end{cases}
\end{equation}
for some suitable region $\cal{D}$, such that it avoids the
interaction of the ``real'' cells with the spurious replicas, while
maintaining all interactions between points in the real cells. In
order to achieve this goal, it will be necessary to increase the size
of the original unit cell.

For finite systems (2D/0D) the cutoff region can be conveniently
defined as $r < R$, for some cutoff radius $R$. We can then easily
perform the Fourier integral in polar coordinates:
\begin{align}
\tilde v^{\rm 0D}(G)&=\int_0^R \mathrm{d}r \int_0^{2\pi}\mathrm{d}\phi
\exp(iGr\cos\phi) \nonumber \\
=2\pi\int_0^R\mathrm{d}rJ_0(Gr) 
&=(2\pi R)\ _{1}F_2\left(\frac{1}{2};1,\frac{3}{2};-\frac{G^2R^2}{4}\right),
\end{align}
where $J_0$ is the Bessel function of the first kind, and $_{1}F_2$ is
the generalized hypergeometric function.
Note that the $G=0$ case is finite and continuous, and poses no difficulty:
\mbox{$
  \tilde{v}^{\rm 0D}(\mathbf{G} = \mathbf{0})=2\pi R.
$}
The value of $R$ must be sufficient to contain all possible interactions
in the original unit cell where the charge is contained; if we imagine it to
be a square of radius $L$, then $R=\sqrt{2}L$. If we now enlarge this cell
(padding the density with zeros) to $L'=(1+\sqrt{2})L$, the spurious replicas
will not interact thanks to the interaction cutoff, and the Hartree integration
will be exact.

The 2D/1D case is more subtle. We assume the charge density to be
contained in a strip defined by $|y|<R/2$: the system is a chain of
unit cells along the $x$ axis.  We define the cutoff region $\cal{D}$
as $|y|<R$ -- and the unit cell is also enlarged in the $y$ direction
to $|y|<R$. It is easy to see that in this manner, the ``ghost''
replicas do not interact with the original chain, but the real
interactions are preserved. The reciprocal space expression for the
truncated Coulomb potential is:
\begin{align}\label{1Dcutoff}
  \tilde v^{\rm 1D}(G_x, G_y)&=\int_{-\infty}^{+\infty}\mathrm{d}x
  \int_{-R}^{+R}\mathrm{d}y
  \frac{\mathrm{e}^{i(G_x x+G_y y)}}{\sqrt{x^2+y^2}} \\
  &=4\int_{0}^R\mathrm{d}y\cos(G_y y)K_0(|G_x y|),
\end{align}
where $K_0$ is the modified Bessel function of the second kind.
However, in this case, since
$
  \lim_{G_x\to 0^+}K_0(G_x y)=+\infty
$,
the integral is undefined on the whole line $G_x = 0$. 

In the fully periodic (2D/2D) case, we also have a singular point at
$\mathbf{G}=\mathbf{0}$ [see Eq.~(\ref{eq:2d-2d})]. If we assume charge
neutrality, this singularity poses no problem. In
Eq.~(\ref{defHartreeReciproc}) the term $v(\mathbf{G}=\mathbf{0})$ multiplies
$n(\mathbf{G}=\mathbf{0})$, which is zero because this is in fact
the charge neutrality condition. In the 2D/1D case, we must estimate how these divergent terms affect Eq.~(\ref{defHartreeReciproc}), or, more
precisely, the back Fourier transform
\begin{equation}
\label{eq:backtransform}
V(\mathbf{r}) = \int {\rm d}^2G e^{-i \mathbf{G}\cdot\mathbf{r}} n(\mathbf{G})v(\mathbf{G})\,.
\end{equation}
In order to see how the infinities appear, we consider the integral in
Eq.~(\ref{1Dcutoff}) for $G_x=0$, but using first a finite integration domain
also in the $x$ direction, $-h < x < h$. This integral is
convergent. If we perform the integration and retain only the terms
that do not vanish in the limit $h \to \infty$ we obtain
\begin{align}
\tilde v^{\rm 1D}(0,G_y)&\approx 4 \log(2h) \sin (G_y R) / G_y \nonumber \\ 
&-4\int_0^R\mathrm{d}y\cos(G_y y)\log(y).
\end{align}
The first term, which we call $v^{\infty}(0,G_y)$, diverges as
$h\to\infty$. However, it can easily be seen that it can be ignored if
we assume charge neutrality. We perform the $G_y$ integration in
Eq.~(\ref{eq:backtransform}) for $G_x=0$, considering only the
$v^{\infty}(0,G_y)$ term
\begin{align}
\int {\rm d}G_y v^{\infty}(0, G_y) n(0, G_y) e^{-i G_y y} = \nonumber \\
4\log(2h) \; \int {\rm d}x' \int_{-R/2}^{R/2} {\rm d}y'   n(x', y')
\int {\rm d}G_y \frac{\sin (Gy R)}{G_y} e^{-i G_y (y-y')} \,.
\end{align}
Now we have:
\begin{equation}
\int {\rm d}G_y  \frac{\sin(Gy R)}{G_y} e^{-i G_y (y-y')} = 
  \begin{cases}
    \pi/2 &    \text{if $|y'-y| < R$} \\
    0&          \text{otherwise.}
  \end{cases}
\end{equation}
In the integral $|y'|<R/2$, since the charge is contained in that region. This is the region
of interest, and therefore we are also interested in looking at the potential only
for $|y|<R/2$. As a consequence, $|y'-y|<R$, and we can conclude
\begin{align}
\int {\rm d}G_y v^{\infty}(0, G_y) n(0, G_y) e^{-i G_y y} = 
\nonumber \\
2\pi\log(2h) \; \int {\rm d}x' \int_{-R/2}^{R/2} {\rm d}y'   n(x', y')\,.
\end{align}
This is obviously zero if we assume charge neutrality. Therefore
we have prooved that we can safely ignore the diverging terms, and retain only the regular ones.
With this in mind, the case of $\mathbf{G}=\mathbf{0}$ is now trivial:
$
  \tilde v^{\rm 1D}(0,0)=-4R(\log R -1).
$

\section{Results}

To test the cutoff method, we consider 1D QD arrays 
similar to those in Ref.~[\onlinecite{karkkainen}]. Each rectangular unit
cell contains two QDs, and each QD has $N$ electrons
bound by a Gaussian positive background charge density. 
The total Gaussian background charge density has the form
\begin{equation}
n_B({\mathbf r}) = \frac{1}{\pi r_s^2}\sum_{\mathbf
  R}\exp\left[-\frac{({\mathbf r}-{\mathbf R})^2}{N r_s^2}\right],
\end{equation}
where ${\mathbf r}=(x,y)$, ${\mathbf R}=(n a_x,0)$ with $n=0,1,2,\ldots$, and
$r_s=2$ is the average density at the center of the QD.
Note the use of eff. a.u. throughout.~\cite{units}

We solve the Kohn-Sham equations within spin-DFT on a 2D
grid with and without the cutoff method described above. 
In the former case, the system is periodic only in {\em x} direction, 
whereas in the latter case it is periodic in both {\em x} and {\em y}
directions. For the exchange and correlation we use the
2D local-spin density approximation (LSDA) with the
parametrization of the correlation by Attaccalite 
{\em et al}.~\cite{attaccalite} This parametrization 
has been shown to be more accurate than the form of 
Tanatar and Ceperley~\cite{tanatar} in the partially 
spin-polarized regime.~\cite{LDA_testing}
All the numerical calculations are done using the
{\tt octopus}~\cite{octopus} code.

\begin{figure}
\includegraphics[width=0.8\columnwidth]{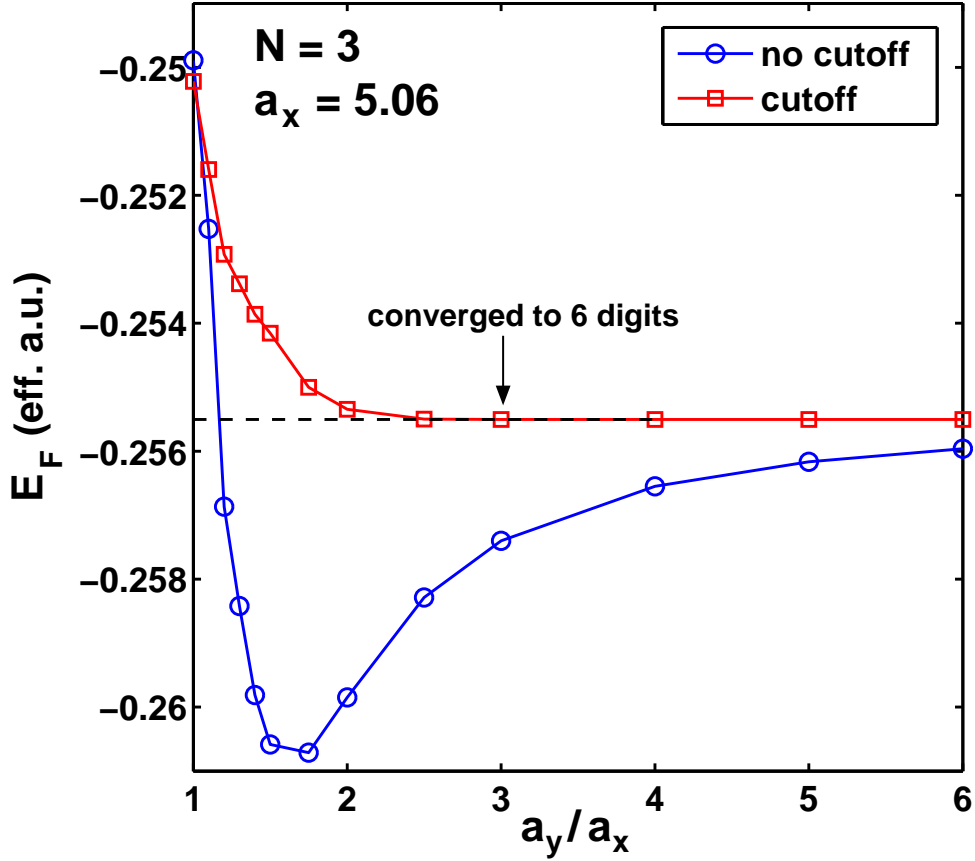}
\caption{(Color online) Fermi energy of a quantum-dot chain with three electrons per dot 
  as a function of the supercell size perpendicular to the chain.}
\label{fig:fermi_energy}
\end{figure}

In Fig.~\ref{fig:fermi_energy} 
we show the Fermi energy $E_F$ of a QD 
chain with three electrons per dot as a function of
the lattice constant $a_y$ perpendicular
to the chain.  The lattice constant
in the {\em x} direction is fixed to $a_x=5.06$. 
Using the cutoff method leads to a very fast convergence, so that
at $a_y=3a_x$ the Fermi energy is converged to six digits. 
In contrast, without cutoff, i.e., the system being 
periodic in both directions, a much larger supercell
is needed in order to achieve comparable accuracy. This is due to the long-range
Coulomb interaction between parallel QD chains.
Therefore, for accurate calculations the 
cutoff scheme is essential in reducing the cell size and
thus the computational cost.

Next we focus our attention to the physical effects
caused by {\em multiple} parallel QD chains in comparison with
a {\em single} QD chain. In the former case no cutoff is used
so that for $a_y=a_x$ we have parallel chains located
at $y=na_x,\; n=\pm 1,\,\pm 2,\ldots$. In the latter case
(single chain) we use the cutoff method with $a_y=3a_x$ to 
guarantee a high precision (see Fig.~\ref{fig:fermi_energy}).

\begin{figure}
\includegraphics[width=0.8\columnwidth]{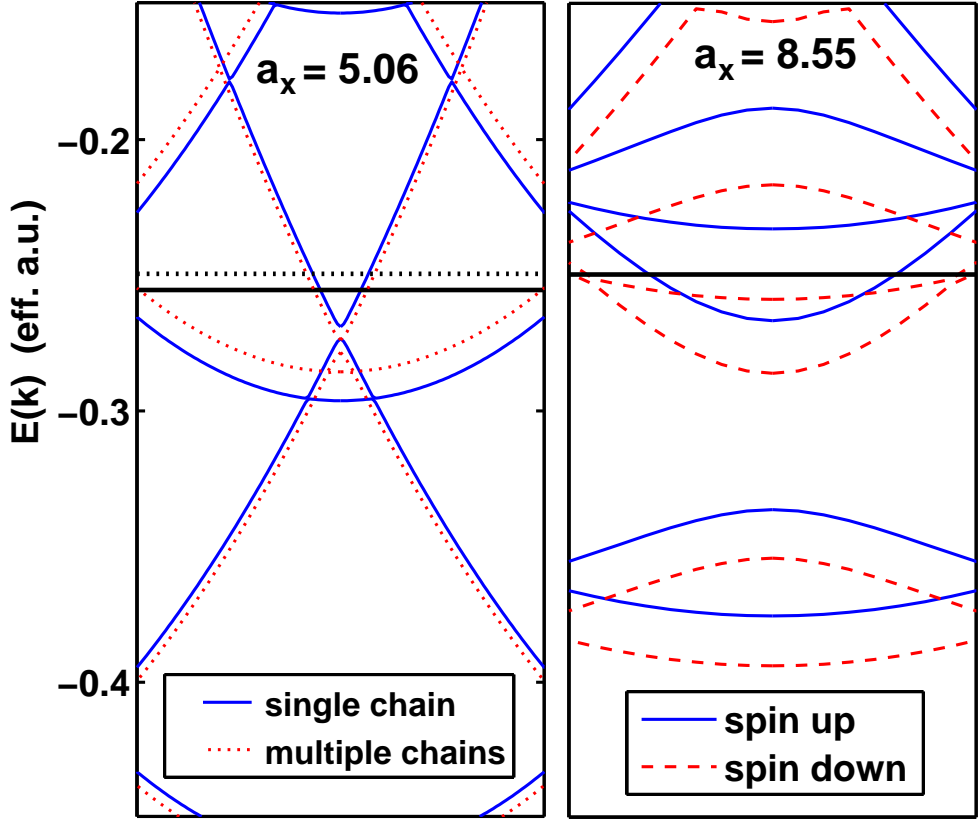}
\caption{(Color online) Band structure at lattice constants $a_x=5.06$ (left) and
$a_x=8.55$ (right) for a quantum-dot chain with three electrons in
each dot. The dotted lines in the left panel correspond to the
result with multiple parallel chains at $y=na_x=na_y,\; n=\pm
1,\,\pm 2,\ldots$. The straight lines correspond to the Fermi energies.}
\label{fig:n3_bands}
\end{figure}

In Fig.~\ref{fig:n3_bands}
we plot the band structure for a QD chain with three electrons per
dot at $a_x=5.06$ (left panel), corresponding to the system considered
in Fig.~\ref{fig:fermi_energy}. 
We find that the presence of parallel chains
leads to a rather minor effect on the bands. 
The qualitative shape
of the structure is very similar, and the shift in the bands due to
the interchain effects is of the same order of magnitude as the shift in
$E_F$. 

Fig.~\ref{fig:n3_bands} can be also directly compared 
with the results of Ref.~[\onlinecite{karkkainen}].
Overall, we find an excellent agreement, also in the case 
of a larger lattice constant $a_x=8.55$ (right
panel). Here the spin degeneracy is lifted due to the exchange effect
which leads to a magnetic ground state.~\cite{karkkainen}

\begin{figure}
\includegraphics[width=0.8\columnwidth]{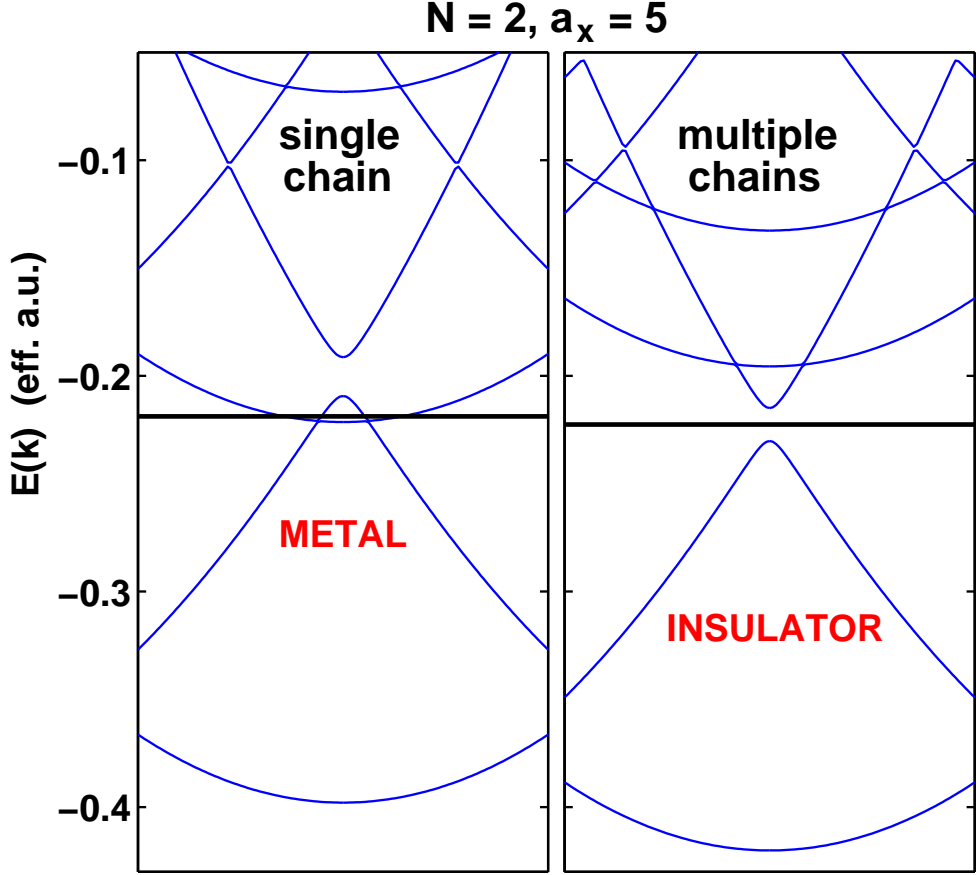}
\caption{(Color online) Band structure at a lattice constant $a_x=5$ for quantum-dot
chains with two electrons in each dot. The left and right panels
show the results for a single and multiple chains,
leading to to metallic and insulating band structures, respectively.
In the latter case, the parallel chains are located at $y=na_x=na_y,\; n=\pm
1,\,\pm 2,\ldots$.}
\label{fig:n2_bands}
\end{figure}

Next we consider the interchain effects on phase transitions
in QD chains. As noted by K\"arkk\"ainen {\em et al.},~\cite{karkkainen}
the chains have a rich phase diagram with respect to the
electron number $N$ and the lattice constant $a_x$. In general,
increasing $a_x$ leads to insulating behavior and/or magnetism, i.e.,
spin polarization, depending on $N$. In Fig.~\ref{fig:n2_bands}
we show the band structure at $a_x=5$ for two electrons per dot,
calculated with the cutoff procedure for a single chain and 
without the cutoff for multiple parallel chains located 
at $y=na_x=na_y,\; n=\pm 1,\,\pm 2,\ldots$.
The single QD chain is clearly a metal with bands crossing
the Fermi level. In contrast the presence of periodic replicas of the chain 
opens up a gap
across the Fermi level so that the system becomes an insulator.
The physical origin of the effect is the increased localization 
of single QDs due to Coulomb repulsion between the replicas 
of the chain. Similarly, the single QD chain (left panel
in Fig.~\ref{fig:n2_bands}) becomes an insulator
at around $a_x=6$, when the dots are more isolated.

Finally, we have also examined the interchain effects on the magnetism
of QD wires. Generally, the spin polarization is relatively stable:
For $N=1\ldots 5$ we find similar spin-polarization as a function of
the lattice constant regardless of the presence of parallel QD
chains. In all cases the magnetization is in a qualitative agreement
with the results in Ref.~[\onlinecite{karkkainen}], minor
differences arising from the use of the LSDA parametrization by
Attaccalite {\em et al}. instead of the one by Tanatar and
Ceperley.~\cite{tanatar} However, further studies are required
regarding the validity of the LSDA in the high-correlation regime
and/or in the presence of magnetic fields.  These would
also be ideal systems to test the 2D density functionals recently
developed.\cite{x2D,c2D}

\section{Summary}

The reciprocal space is the natural venue for treating periodic systems. If the
periodicity is not complete (the system is not periodic in all the space
dimensions) or even absent, the computations require the use of a large
supercell in order to avoid the spurious interactions due to ``ghost'' system
replicas; this is specially manifested in the computation of the Coulomb, or
Hartree, integral, which is a trivial multiplication if we use plane waves. For
3D systems, it has been shown that the use of a screened Coulomb interaction
greatly improves the accuracy in the calculation of ground states quantities,
and substantially simplifies the evaluation of excited-state properties of
reduced-periodicity systems. In this work we have shown that the same
ideas can be applied to the case of the 2D electron gas; we have provided the relevant
formulae for finite systems in 2D, and for systems that are periodic in only
one dimension. 

Moreover, the corrective cutoffs are exact and rather straightforward to apply.
We have demonstrated their effectiveness by computing band structures and Fermi
energies of one-dimensional periodic arrays of quantum dots formed by
the confinement of two-dimensional
electron gas. We expect that our technique will be of great interest for
studying these novel systems of ``artificial molecules'' and ``crystals'',
although the ubiquity of the problem we have faced will probably make our
technique relevant in other fields, too.

\begin{acknowledgments}
This work was supported by the EC Network of Excellence
NANOQUANTA (NMP4-CT-2004-500198), the 
Deutsche Forschungsgemeinschaft within the SFB 658,
and the Academy of Finland.
\end{acknowledgments}


\begin{thebibliography}{99}
\bibitem{units} All the formulae given here are the same 
in effective atomic units and in Hartree atomic units.
Effective atomic units
are convenient to account for the background material of
quantum-dot systems. The lengths and energies are then given in
$a^*_0 =(\epsilon/m^*)a_0$ and ${\rm Ha}^*  = (m^*/\epsilon^2)\,{\rm
  Ha}$, respectively. Here $m^*$ is the effective mass and
 $\epsilon$ is the dielectric constant of the background material.
\end{thebibliography}
\end{document}